\documentclass[11pt]{article}
\usepackage{graphicx}
\usepackage{amssymb}
\usepackage{epstopdf}

\title{\bf A New Lower Bound for \boldmath{$A(17,6,6)$}}

\author{{\em Yeow Meng Chee}\\
Interactive Digital Media Program Office\\
Media Development Authority\\
140 Hill Street\\
Singapore 179369\\
and \\
Division of Mathematical Sciences\\
School of Physical and Mathematical Sciences\\
Nanyang Technological University\\
Singapore 637616
}

\date{}

\begin{document}

\maketitle

\abstract{
We construct a record-breaking binary code of length $17$, minimal distance $6$,
constant weight $6$, and containing $113$ codewords.
}

\section{Introduction}

Let $A(n,d,w)$ denote the maximum possible number of codewords in a binary code of length $n$, minimal distance $d$ and constant weight $w$. The Nordstrom-Robinson code ${\cal N}_{16}$ of length $16$, minimal distance $6$, and containing $256$ codewords has weight enumerator $1+112x^6+30x^8+112x^{10}+x^{16}$. Hence, taking all the codewords of weight $6$ in
${\cal N}_{16}$ gives a constant weight code that shows $A(16,6,6)\geq 112$. Since
$A(17,6,6)\geq A(16,6,6)$, we also have $A(17,6,6)\geq 112$. This is in fact the best lower bound on $A(17,6,6)$ known \cite{Brouweretal90}.

In this note, we give the first improvement on the lower bound for $A(17,6,6)$ since that implied by the 1967 result of Nordstrom and Robinson \cite{NordstromRobinson67}. We exhibit a new binary code $\cal C$ of length $17$, minimal distance $6$, constant weight $6$, and containing $113$ codewords, showing $A(17,6,6)\geq 113$. Our code has no particular structure (its automorphism group is trivial) and is obtained through a combination of search techniques involving simulated annealing \cite{NurmelaOstergard93}, length-reduction \cite{Awetal03}, and local optimization.

The support ${\rm supp}(x)$ of a codeword $x=(x_1,\ldots,x_n)$ is the set of indices of its non-zero coordinates, that is, ${\rm supp}(x)=\{i\mid x_i\not= 0\}$. The supports of the codewords in $\cal C$ are listed in the next section.

\section{The Code}

\begin{center}
\begin{tabular}{|c|c|c|}
\hline
0 1 2 3 6 15 & 0 1 2 4 11 16 & 0 1 2 7 8 9 \\
0 1 2 10 12 13 & 0 1 3 4 8 10 & 0 1 3 5 7 12 \\
0 1 3 9 13 16 & 0 1 4 6 7 13 & 0 1 5 6 10 16 \\
0 1 5 8 11 13 & 0 1 6 9 11 12 & 0 1 7 10 11 15 \\
0 1 8 12 14 15 & 0 2 3 4 9 12 & 0 2 3 5 8 16 \\
0 2 3 7 11 13 & 0 2 4 5 7 10 & 0 2 4 8 13 15 \\
0 2 5 6 9 13 & 0 2 5 11 14 15 & 0 2 6 7 12 16\\
0 2 6 8 10 11 & 0 3 4 5 6 11 & 0 3 4 7 14 16 \\
0 3 5 10 13 15 & 0 3 6 7 9 10 & 0 3 6 8 12 13 \\
0 3 8 9 11 15 & 0 3 10 11 12 14 & 0 4 5 12 13 14 \\
0 4 6 8 9 16 & 0 4 6 10 12 15 & 0 4 7 8 11 12 \\
0 4 9 10 11 13 & 0 5 6 7 8 15 & 0 5 8 9 10 14 \\
0 5 9 12 15 16 & 0 6 11 13 14 16 & 0 7 8 10 13 16 \\
0 7 9 13 14 15 & 1 2 3 4 5 13 & 1 2 3 7 10 14 \\
1 2 3 8 11 12 & 1 2 4 6 9 10 & 1 2 4 7 12 15 \\
1 2 5 6 7 11 & 1 2 5 8 10 15 & 1 2 5 12 14 16 \\
1 2 6 8 13 16 & 1 2 9 11 13 15 & 1 3 4 6 12 16 \\
1 3 4 7 9 11 & 1 3 5 6 8 14 &1 3 5 11 15 16 \\
1 3 6 10 11 13 & 1 3 7 8 13 15 & 1 3 9 10 12 15 \\
1 4 5 7 8 16 & 1 4 5 9 14 15 & 1 4 5 10 11 12 \\
1 4 6 8 11 15 & 1 4 8 9 12 13 & 1 4 10 13 14 16 \\
1 5 6 12 13 15 & 1 5 7 9 10 13 & 1 6 7 8 10 12 \\
1 6 7 9 15 16 & 1 7 11 12 13 16 & 1 8 9 10 11 16 \\
2 3 4 6 7 8 & 2 3 4 10 11 15 & 2 3 5 6 10 12 \\
2 3 5 7 9 15 & 2 3 6 9 11 16 & 2 3 8 9 10 13 \\
2 3 12 13 15 16 & 2 4 5 6 15 16 & 2 4 5 8 9 11 \\
2 4 6 11 12 13 & 2 4 7 9 13 16 & 2 4 8 10 12 14 \\
2 5 7 8 12 13 & 2 5 10 11 13 16 & 2 6 7 10 13 15 \\
2 6 8 9 12 15 & 2 7 8 11 15 16 & 2 7 9 10 11 12 \\
2 9 10 14 15 16 & 3 4 5 8 12 15 & 3 4 5 9 10 16 \\
3 4 6 13 14 15 & 3 4 7 10 12 13 & 3 4 8 11 13 16 \\
3 5 6 7 13 16 & 3 5 7 8 10 11 & 3 5 9 11 12 13 \\
3 6 7 11 12 15 & 3 6 8 10 15 16 & 3 7 8 9 12 16 \\
4 5 6 7 9 12 & 4 5 6 8 10 13 & 4 5 7 11 13 15 \\
4 6 7 10 11 14 & 4 7 8 9 10 15 & 4 11 12 14 15 16 \\
5 6 8 11 12 16 & 5 6 9 10 11 15 & 5 7 9 11 14 16 \\
5 7 10 12 14 15 & 5 8 13 14 15 16 & 6 7 8 9 11 13 \\
6 9 10 12 13 16 & 8 10 11 12 13 15 & \\
\hline
\end{tabular}
\end{center}


\thebibliography{X}

\bibitem{Awetal03}
H. K. Aw, Y. M. Chee and A. C. H. Ling, "Six new constant weight binary codes", {\em Ars Combin.}, {\bf 67} (2003), 313--318.

\bibitem{Brouweretal90}
A. E. Brouwer, J. B. Shearer, N. J. A. Sloane and W. D. Smith, "A new table of constant weight codes", {\em IEEE Trans. Inform. Theory}, {\bf 36} (1990), 1334--1380.

\bibitem{NordstromRobinson67}
A. W. Nordstrom and J. P. Robinson, "An optimum nonlinear code", {\em Inform. Control}, {\bf 11} (1967), 613--616.

\bibitem{NurmelaOstergard93}
K. J. Nurmela and P. R. J. \"{O}sterg{\aa}rd, {\em Upper bounds for covering designs by
simulated annealing}, Congr. Numer. {bf 96} (1993), 93--111.
 \end{document}